# Hydrodynamic fluctuations of a liquid with anisotropic molecules

A. V. Zatovsky[1,2] and A. V. Zvelindovsky[3]

[1] Colloid and Interface Science group, LIC, Leiden University, P.O. Box 9502, 2300 RA Leiden, The Netherlands

[2] Department of Theoretical Physics, Odessa National University, Dvoryanskaya 2, 65026, Odessa, Ukraine

[3] Soft Condensed Matter group, LIC, Leiden University, P.O. Box 9502, 2300 RA Leiden, The Netherlands

**Abstract**
A general theory of hydrodynamic fluctuations of a liquid with anisotropic molecules, in the presence of steady simple shear, has been proposed.

**1. Introduction**

Thermal motion of molecules of liquids manifests itself in many physical experiments. One of the clearest example is scattering or absorption of radiation by liquids. The theoretical description of these experiments is based on the calculation of the time dependent correlation functions (CFs) of molecular variables. An exact calculation is not possible due to the difficulty of describing systems of many interacting particles. Therefore, the analysis of experimental data is often based upon a model.

The simplest, but important, descriptors of translational and rotational movement in liquids are CFs of translational and rotational velocities $(\vec{v}, \vec{\omega})$ of molecules

$$\varphi_v(t) = \langle \vec{v}(t)\, \vec{v}(0) \rangle, \quad \varphi_\omega(t) = \langle \vec{\omega}(t)\, \vec{\omega}(0) \rangle , \qquad (1.1)$$

and the CF of spherical harmonics of Euler angles, determining the orientation of the molecular reference frame with respect to a laboratory one:

$$\Psi_{lm}(t) = \langle Y_{lm}(t)\, Y_{lm}^*(0) \rangle , \qquad (1.2)$$

where is averaged over the distribution of molecular variables. These functions are directly related to selfdiffusion and incoherent neutron scattering, dielectric relaxation, NMR, light scattering, etc.

The behaviour of the functions (1.1-1.2) for various representations of intermolecular interactions was thoroughly studied by molecular dynamics. In the case of simple liquids and liquid metals close to the triple point, the function $\varphi_v(t)$ oscillates, and it decays as $t^{-3/2}$ for large $t$. According to molecular dynamics, the CF of rotational velocity demonstrates oscillations as well, and the CF of $Y_{lm}(t)$ usually decays monotonically.

Molecular dynamics investigations support the idea that molecular motions in liquids is a collective phenomenon. The collective nature of the long-range correlations is clearly demonstrated by the algebraic decay of the time dependent CF of the velocity. This result has stimulated a huge flow of articles studying long living correlations in disordered systems and their manifestation in kinetic processes. The behaviour of the CF of hydrodynamic fields in an inhomogeneous environment, including a detailed description of translational and rotational molecular motions in a wide time and frequency range for the simple liquids and liquids with isotropic molecules, is of special importance.

One of the methods used for studying the thermal fluctuations of local macroscopic variables is based upon solving inhomogeneous linear Langevin equations, that describe the dynamics of fluctuations. These equations can be obtained by linearizing the macroscopic equations of thermodynamically nonequilibrium processes around equilibrium and adding the Gauss fields of fluctuating fluxes, which are δ-correlated in time (as sources of thermal noise). The statistical properties of these sources are determined by the fluctuation-dissipation theorem. Langevin method has been widely applied in the theory of equilibrium thermal fluctuations, for instance electromagnetic and hydrodynamic fluctuations, and in the kinetic theory of gases, taking into account fluctuations of distribution function, etc [1-4].

The theory of nonequilibrium hydrodynamic fluctuations in inhomogeneous stationary states was studied by many authors [4-12]. The reason is that the study of hydrodynamic fluctuations with large correlation radius allows for the analysis of dissipative structures in inhomogeneous flows and the development of hydrodynamic instabilities in these systems [13]. The main problem of the general theory of nonequilibrium hydrodynamic fluctuations is the description of the influence of nonequilibrium conditions on the process of generation of thermal hydrodynamic noise. A further problem is in obtaining the large-scale fluctuations responsible for long wave correlations in the nonequilibrium system. The most detailed description can be found in [4]. In the case of stationary states, the nonequilibrium hydrodynamic fluctuations are responsible for the asymmetry of Brillouin spectra of light scattered by inhomogeneously heated liquid [8,14-17] or by liquid in the presence of shear flow [9,18].

Many papers have been devoted to a simpler problem such as the influence of nonequilibrium fluctuations on Brownian diffusion. Diffusion and friction coefficients of a Brownian particle have been studied in detail in the vicinity of natural convection in a heated liquid [19,20] or in a general Couette flow [21-25]. Dilute suspension of hard spheroidal particles experiencing rotational Brownian motion in a Couette flow is the simplest model of colloids or liquid with anisotropic molecules in an inhomogeneous situation.

A phenomenological description of the thermal motion in a liquid consisting of non-spherical molecules, and a theoretical study of light scattering in such a system, was done by M. Leontovich [26]. He supposed that the liquid can be described by the usual hydrodynamic variables (density, pressure, temperature, velocity) and additionally by an anisotropy tensor, describing the deviation of the axes of anisotropic molecules in a small liquid volume from isotropic distribution. In more than half a century that passed, the theory [26] has been generalized and developed further (see review [27]). The dynamical equations for colloidal systems with a tensor order parameter have been derived and analysed in [28,29].

Anisotropic molecules rotate and have a tensor of inertia. It is natural to introduce a local inertia tensor of a volume of liquid. Curtiss has used such a tensor in the kinetic theory of molecular gases [30]. In nonequilibrium, the tensor depends on the temperature and the pressure. Changes in the latter ones lead to a new equilibrium value of the inertia tensor. As this is not an instant process, the tensor can be taken as an independent state variable, similar to the Leontovich anisotropy tensor. Deviation of this tensor from the equilibrium value is related to mechanical and electrical anisotropy of the liquid [26].

In the present work we obtain time and space dependent CFs of the spectral density of the fluctuations of the inertia tensor for a liquid particle in equilibrium and nonequilibrium conditions in the presence of a simple shear flow. The description is based on a full system of hydrodynamic equations, where in addition to the local velocity $\vec{v}(\vec{r},t)$ we take into account internal momentum $\vec{M}(\vec{r},t)$ and the tensor of inertia $I_{\alpha\beta}(\vec{r},t)$.

## 2 Generalized hydrodynamics and spectral density of the CFs of inertia tensor

The local conservation laws of mass, energy and momentum read

$$\frac{\partial \rho}{\partial t} + div(\rho \vec{v}) = 0 \quad (2.1)$$

$$\frac{\partial E}{\partial t} + div\vec{Q} = 0 \quad (2.2)$$

$$\frac{\partial}{\partial t}\rho v_\alpha + \nabla_\beta(\rho v_\alpha v_\beta) = \nabla_\beta(\sigma_{\alpha\beta} - p\delta_{\alpha\beta}) \quad (2.3)$$

where $\rho$ is density, E is the total energy of the unit volume of liquid, $\vec{Q}$ is the energy flux, $p$ is the pressure and $\sigma_{\alpha\beta}$ is the stress tensor. Conservation of the total moment of momentum $\rho[\vec{r},\vec{v}] + \vec{M}$ can be expressed as

$$\frac{\partial M_\alpha}{\partial t} + \nabla_\beta(v_\beta M_\alpha) = \nabla_\beta g_{\beta\alpha} + \varepsilon_{\alpha\beta\gamma}\sigma_{\gamma\beta} \quad (2.4)$$

where $g_{\alpha\beta}$ is the tensor of flux of the moment of momentum, $\varepsilon_{\alpha\beta\gamma}$ is the unit antisymmetric tensor, and we have eliminated the angular momentum related to translational motion. The total hydrodynamic system of equations includes the equations for the inertia tensor and entropy production

$$\left(\frac{\partial}{\partial t} + \nabla_\gamma v_\gamma\right)\rho I_{\alpha\beta} = \nabla_\gamma P_{\alpha\beta\gamma} + D_{\alpha\beta} \quad (2.5)$$

$$\left(\frac{\partial}{\partial t} + \nabla_\gamma v_\gamma\right)\rho s = \sigma \quad (2.6)$$

The quantities $\vec{Q}, g_{\alpha\beta}, \sigma_{\alpha\beta}, D_{\alpha\beta}, P_{\alpha\beta\gamma}, \sigma$ will be clarified in the remainder. Such a system of equations has been found for gasses in [30] by means of the modified Boltzmann equation. The density of total energy of the liquid reads

$$E = \frac{1}{2}\rho v^2 + \tilde{E} + \vec{M}\,\vec{\Omega} ,$$
(2.7)

where $\tilde{E}$ is the internal energy in the reference frame moving with an element of liquid volume. This reference frame rotates with respect to the initial one with an angular velocity $\vec{\Omega} = \frac{1}{2} rot\,\vec{v}$. The internal energy is a function of the new thermodynamic variables, namely, moment of momentum and inertia tensor

$$\tilde{E} = E_0(\vec{M}, I) - \vec{M}\,\vec{\Omega} .$$
(2.8)

Thus, the main thermodynamic identity is

$$d\tilde{E} = \rho\,T\,ds + \omega\,d\rho + \left(M_\alpha I^{-1}_{\alpha\beta} - \Omega_\beta\right) dM_\beta - M_\alpha d\Omega_\alpha + B\rho\,I_{\alpha\beta} dI_{\alpha\beta} + o(SpI^3) ,$$

where $\omega$ is the enthalpy of unit mass and $B$ is an expansion coefficient. Using (2.8) and (2.1-2.6) one can find in a standard way [31,32] the entropy production

$$\sigma = \kappa_{\alpha\beta}\left(\nabla_\alpha v_\beta - \omega_\gamma \varepsilon_{\alpha\beta\gamma}\right) + g_{\alpha\beta}\nabla_\alpha v_\beta + B\,P_{\alpha\beta\gamma}\nabla_\gamma I_{\alpha\beta} + B\,D_{\alpha\beta} I_{\alpha\beta} ,$$
(2.9)

where

$$\kappa_{\alpha\beta} = \sigma_{\alpha\beta} + \delta_{\alpha\beta}(\tilde{\omega} - \vec{M})\vec{M},\quad \omega_\alpha = I^{-1}_{\alpha\beta} M_\beta / \rho .$$
(2.10)

The Curie theorem and Onsager relations leads to the phenomenological equations

$$\kappa_{\alpha\beta} = \eta\left(\vec{\nabla}\,\vec{v}\right)^s_{\alpha\beta} + \zeta\,\delta_{\alpha\beta}\vec{\nabla}\cdot\vec{v} + \eta_3 \varepsilon_{\alpha\beta\gamma}(\omega_\gamma - \Omega_\gamma) + \eta_{12}\delta I^s_{\alpha\beta} ,$$

$$g_{\alpha\beta} = \mu_1 \nabla_\beta \omega_\alpha,\quad P_{\alpha\beta\gamma} = \mu_2 \nabla_\gamma I_{\alpha\beta} ,$$
(2.11)

$$D_{\alpha\beta} = \eta_{21}\left(\vec{\nabla}\,\vec{v}\right)^s_{\alpha\beta} - \eta_2 \delta I_{\alpha\beta} .$$

Here, the index $s$ stands for the traceless part of symmetric tensor (deviator) and $\delta I_{\alpha\beta}$ is the deviation of the inertia tensor from the equilibrium value. Independent phenomenological coefficients $\eta,\ \zeta,\ \eta_3,\ \mu_1$ of first, second and third (rotational) viscosities and the diffusion coefficient of internal momentum are the same as in [32], and the coefficients $\eta_2,\ \eta_{12},\ \eta_{21},\ \mu_2$ are new ones, due to the introduction of an extra thermodynamic variable. Thus, the generalized hydrodynamics reads

$$\left(\frac{\partial}{\partial t} + \nabla_\sigma v_\sigma\right)\rho\,v_\alpha = -\nabla_\alpha p + \eta\,\Delta v_\alpha + \left(\zeta + \frac{\eta}{3}\right)\nabla_\alpha div\,\vec{v} + \eta_3\,rot_\alpha(\vec{\omega} - \vec{\Omega}) + \eta_{12}\nabla_\beta \delta I^s_{\alpha\beta} ,$$

$$\left(\frac{\partial}{\partial t}+\nabla_\sigma v_\sigma\right)\vec{M}=-\eta_3\left(\vec{\omega}-\vec{\Omega}\right)+\mu_1\Delta\vec{M},$$

(2.12)

$$\left(\frac{\partial}{\partial t}+\nabla_\sigma v_\sigma\right)\rho\, I^س_{\alpha\beta}=\mu_2\Delta I^س_{\alpha\beta}-\eta_2\,\delta I^س_{\alpha\beta}+\eta_{21}(\nabla\vec{v})^س_{\alpha\beta}.$$

The system is closed by the continuity equation, entropy production and the equation of state.

After linearization, this systems allows for the calculation of fluctuations for all hydrodynamic fields, using, for example, the fluctuation-dissipation theorem. We are only interested in the spectral densities of deviators of the inertia tensor. After some simple but cumbersome algebra one obtains

$$\left\langle I^d_{\alpha\beta} I^{d*}_{\gamma\delta}\right\rangle_{\omega k}=\mathrm{Re}\frac{\theta}{M}\left\{2\Delta_{\alpha\beta\gamma\delta}-\frac{k^2}{NM}\frac{b}{2}R_{\alpha\beta\gamma\delta}+i\omega\, J\, N\,\Delta_{\alpha\beta}\Delta_{\gamma\delta}\right\}.$$

(2.13)

Where we have introduced the notations

$$\theta=\frac{k_B T}{8B\,\rho_0\pi^4},\quad b=\frac{\eta_{21}\,\eta_{12}}{2\rho_0^2},$$

$$\Delta_{\alpha\beta}=n_\alpha n_\beta-\frac{1}{3}\delta_{\alpha\beta},\quad \Delta_{\alpha\beta\gamma\delta}=\frac{1}{2}\left(\delta_{\alpha\gamma}\delta_{\beta\delta}+\delta_{\alpha\delta}\delta_{\beta\gamma}\right)-\frac{1}{3}\delta_{\alpha\beta}\delta_{\gamma\delta}$$

$$R_{\alpha\beta\gamma\delta}=\frac{1}{4}\left(n_\alpha n_\delta\delta_{\gamma\beta}+n_\alpha n_\gamma\delta_{\beta\delta}+n_\alpha n_\beta\delta_{\gamma\delta\gamma}+n_\beta n_\gamma\delta_{\alpha\delta}\right)-$$

(2.14)

$$-\frac{1}{3}\left(n_\alpha n_\beta\delta_{\gamma\delta}+n_\gamma n_\delta\delta_{\alpha\beta}\right)+\frac{1}{9}\delta_{\alpha\beta}\delta_{\gamma\delta},\quad n_\alpha=\frac{k_\alpha}{k},$$

$$M=-i\omega+\frac{k^2\mu_2}{\rho_0}+\frac{\eta_2}{\rho_0},\quad N=-i\omega+\frac{k^2\eta}{\rho_0}+\frac{k^2 b}{M},$$

(2.15)

$$J^{-1}=\omega^2-c^2 k^2+i\omega\, k^2\left(\frac{\zeta}{\rho_0}+\frac{4}{3}\nu\right)+\frac{2b}{M},$$

where $\rho_0$ is an equilibrium density and $c$ is the isothermic sound velocity. As it is shown in [32, 33] the moment of momentum equilibrates fast and therefore we have omitted terms describing the influence of change of moment of momentum on fluctuations of the inertia tensor.

We calculate the CF of components of the inertia tensor of a Lagrange fluid particle as

$$\Psi^L_{\alpha\beta\gamma\delta}(t)=\left\langle I^d_{\alpha\beta}(\vec{a}(t),t)\, I^d_{\alpha\beta}(\vec{a}(0),0)\right\rangle.$$

(2.16)

Instead of the Cartesian components of the symmetric tensor of second rank, one can introduce its spherical components

$$I_{\pm 2} = \frac{1}{4}(I_{xx} - I_{yy} \pm 2iI_{xy}), \quad I_{\pm 1} = \frac{1}{2}(I_{xz} \pm iI_{yz}), \quad I_0 = -(I_{xx} + I_{yy})$$

The CF of them

$$\Psi_m^I(t) = \langle I_m(\vec{a}(t),t) \, I_m^*(\vec{a}(0),0) \rangle \tag{2.17}$$

is a good approximation for the molecular CF (1.2). Here, the time $t$ is taken in the hydrodynamic range. Thus,

$$\Psi_{2m}(t) \approx \Psi_m^I(t) \tag{2.18}$$

We can eliminate the weak time dependency of the vector argument in (2.16) by using a method of [34] (see also [35]). As a result the Lagrange CF of the inertia tensor can be written in a form of a series of Euler CF and the mean square displacement, $G(t)$, of a liquid particle

$$\Psi^L_{\alpha\beta\gamma\delta}(t) = \sum_{n=0}^{\infty} \frac{1}{n!}\left(\frac{\Gamma(t)}{6}\right)^n \langle I^d_{\alpha\beta}(\vec{r},t) \, \Delta^n I^d_{\gamma\delta}(\vec{r},0) \rangle \tag{2.19}$$

The Euler CF of thermal fluctuations of the inertia tensor can be expressed via spectral densities of intensity of fluctuations

$$\langle I^d_{\alpha\beta}(0,0) \, I^d_{\gamma\delta}(\vec{r},t) \rangle = \int d\vec{k} \, d\omega \, \langle I^d_{\alpha\beta} I^{d*}_{\gamma\delta} \rangle_{\omega,\vec{k}} \, e^{i(\vec{k}\cdot\vec{r} - \omega t)} \tag{2.20}$$

Substituting one into another, equations (2.20, 2.19) give us after summation

$$\Psi^L_{\alpha\beta\gamma\delta}(t) = \int d\vec{k} \, d\omega \, \langle I^d_{\alpha\beta} I^{d*}_{\gamma\delta} \rangle_{\omega,\vec{k}} \, e^{-i\omega t - k^2 \Gamma(t)/6} \tag{2.21}$$

Thus, the Lagrange CFs are expressed via spectral densities of fluctuations of the Euler fields.

The spectral density of fluctuations (2.13) consist of three terms. The first one contributes to the time dependent CF $\Psi^L_{\alpha\beta\gamma\delta}(t)$ as follows

$$\Psi^{(1)}_{\alpha\beta\gamma\delta}(t) = \frac{k_B T \Delta_{\alpha\beta\gamma\delta}}{\rho_0 B (4\pi(Dt + \Gamma(t)/6))^{3/2}} e^{-t/\tau}, \tag{2.22}$$

where $\tau = \dfrac{\rho_0}{\eta_2}$ is the relaxation time and $D = \dfrac{\mu_2}{\rho_0}$ is the diffusion coefficient of inertia tensor. The second contribution can be written after integration

$$\Psi^{(2)}_{\alpha\beta\gamma\delta}(t) = \frac{3\pi^{3/2} \theta b \Delta_{\alpha\beta\gamma\delta}}{2(\nu - D)^{5/2} \sqrt{t}} \int_0^1 \frac{x e^{-xt/\tau}}{(\lambda - x)^{5/2}} \Phi\left(\frac{5}{2}, 2, -\frac{bt}{\nu - \mu}\frac{x(1-x)}{\lambda - x}\right) dx \tag{2.23}$$

where $\Phi(n,m,x)$ is a degenerate hyperbolic function [36], and $\lambda = (v + \Gamma(t)/6)/(v - D)$, $\lambda > 1$.

These two contributions lead to incorrect values of the CFs at small times, since the value of Y(0) does not exist due to a bad determined Euler CF at coinciding space and time arguments [34]. However, we are only interested in the asymptotic behaviour of the Lagrange CF of inertia tensor at $t \to \infty$. It is a good asymptotic approximation for the molecular CF consisting of slowly changing second rank tensors due to rotational motion of molecules of liquid. The contribution (2.22) into the Lagrange CF $\Psi^L_{\alpha\beta\gamma\delta}(t)$ at $t \to \infty$ is exponentially small. One can show that the third contribution arising from (2.13) is exponentially small as well. In addition to sound modes of the longitudinal Euler motions [34], these contributions include relaxation of the inertia tensor. The expression (2.23) in addition to exponentially decaying terms has slowly decaying algebraic terms. The main one is

$$\Psi^L_{\alpha\beta\gamma\delta}(t) = \frac{3k_B T b \tau^2 \Delta_{\alpha\beta\gamma\delta}}{16\rho_0 B \pi \left( t(v + D_L + b\tau) \right)^{5/2}} \left[1 + o\left(\frac{\tau}{t}\right)\right], \quad (2.24)$$

where $D_L$ is the diffusion coefficient of Lagrange particle [34,35].

The algebraic asymptote of the CF of the inertia tensor is determined by the interplay of fluctuations of the transpose Euler motions and fluctuations of the tensor itself. This process is determined by the kinetic coefficients $\eta_{12}, \eta_{21}$ which are incorporated into the term *b* in (2.14). Neglecting this effect leads to a decoupling of the equation of motion for the inertia tensor from other hydrodynamic equations. Then, the Lagrange CF is fully determined by (2.24).

## 3. Time evolution of fluctuations in an anisotropic liquid in a stationary flow

Let us consider the CFs of Euler fields in the presence of stationary flow. We will be interested only in the fields of density, velocity and inertia tensor. In the co-moving reference frame the Langevin equations for the small fluctuating contributions to stationary fields have the form

$$\rho = \rho_0 + \rho', \quad \vec{v} = \vec{u} + \vec{v}', \quad I^d_{\alpha\beta} = \tau v_{21}\left(\frac{\partial u_\alpha}{\partial x_\beta}\right)^s + Q_{\alpha\beta}, \quad (3.1)$$

$$\frac{\partial \rho'}{\partial t} + div(\rho_0 \vec{v}) + div(\rho' \vec{u}) = 0,$$

$$\left(\frac{\partial}{\partial t} + u_\sigma \nabla_\sigma\right) v'_\alpha + v'_\sigma \nabla_\sigma u_\alpha = -\frac{1}{\rho_0}\nabla_\alpha p + v \Delta v'_\alpha + \left(\frac{\xi}{\rho_0} + \frac{v}{3}\right)\nabla_\alpha div\, \vec{v}' + v_{12}\nabla_\sigma Q_{\alpha\sigma} + \frac{1}{\rho_0} f_\alpha(\vec{r},t),$$

$$\left(\frac{\partial}{\partial t} + u_\sigma \nabla_\sigma\right) Q_{\alpha\beta} = D\Delta Q_{\alpha\beta} - \frac{1}{\tau} Q_{\alpha\beta} + v_{21}\left(\nabla \vec{v}'\right)^s_{\alpha\beta} + \frac{1}{\rho_0} F_{\alpha\beta}(\vec{r},t). \quad (3.2)$$

Here $f_\alpha$ and $F_{\alpha\beta}$ are the spontaneous sources of fluctuations.

In what follows, we will study the fluctuations in the presence of simple steady shear

$u_\alpha = \Gamma_{\alpha\sigma} x_\sigma$ ($\Gamma_{\sigma\sigma} = 0$). Let $\varphi_a(\vec{r},t)$, $a=1,2,…,8$, be an eight-fold vector $\{\rho', \vec{v}', Q_{\alpha\beta}\}$ consisting of the density, velocity, and five components of inertia tensor. Then (3.1, 3.2) can be written as

$$\frac{\partial \varphi_a(\vec{r},t)}{\partial t} + L_{ab}\, \varphi_b(\vec{r},t) = f_a(\vec{r},t), \tag{3.3}$$

where $L_{ab}$ is matrix of linear differential operators, which can be found from (3.2) after eliminating the pressure. Let us consider the CFs

$$\Phi_{ab}(\vec{r},\vec{r}',t-t') = \langle \varphi_a(\vec{r},t)\varphi_b(\vec{r}',t') \rangle, \tag{3.4}$$

where is averaged over nonequilibrium fluctuations. The Langevin equations allows for determining the CFs (3.4) via correlators of spontaneous sources [37]. Such a method has been used in [12] to study the standard hydrodynamic fluctuations in the presence of shear. We will follow another route. Taking into account that spontaneous sources and fluctuating hydrodynamic fields at different times do not correlate, we have for the CF (3.4) the following linear homogeneous equations

$$\frac{\partial}{\partial t}\Phi_{ac}(\vec{r},\vec{r}',t-t') + L_{ab}\Phi_{bc}(\vec{r},\vec{r}',t-t') = 0, \tag{3.5}$$

which should be accompanied by the initial conditions

$$\Phi_{ac}(\vec{r},\vec{r}',0) = \langle \varphi_a(\vec{r},t)\varphi_c(\vec{r}',t) \rangle. \tag{3.6}$$

Thus, we have a boundary problem with initial conditions. Equilibrium thermodynamic fluctuations in an unconfined volume are well known, and in case of generalized hydrodynamics, described in the previous chapter. First, the spatial correlations of nonequilibrium fluctuations have been studied by Hinton [5], and later they have been used in [8] for simple liquids. For instance, the stationary velocity correlator in Fourier space is [5]

$$\langle v_\alpha(k,t)v_\beta^*(k,t) \rangle = \frac{1}{\rho_0}\left[k_B T\, \delta_{\alpha\beta} - m\nu\left(\frac{\partial u_\alpha}{\partial x_\beta} + \frac{\partial u_\beta}{\partial x_\alpha} - \tfrac{2}{3}\delta_{\alpha\beta}\,\mathrm{div}\,\vec{u}\right) - \frac{m\xi}{\rho_0}\delta_{\alpha\beta}\,\mathrm{div}\,\vec{u}\right], \tag{3.7}$$

where *m* is the gas molecule mass. Further, the analysis of nonequilibrium fluctuations in a gas, based on modified Boltzmann equation [4], has shown that Hinton included only the thermal noise due to intermolecular interactions in physically infinitely small volumes, so-called short-range fluctuations. Taking into account the long-range fluctuations at weak nonequilibrium conditions gives rise to the additional contribution to equation (3.7), which in the linear gradient approximation is

$$\delta\langle v_\alpha(k,t)v_\beta^*(k,t) \rangle =$$

$$-\frac{k_B T}{2\rho_0}\left[\frac{1}{v_l k^2}n_\alpha n_\beta n_\sigma n_\gamma + \frac{1}{v\, k^2}(\delta_{\alpha\sigma} - n_\alpha n_\sigma)(\delta_{\beta\gamma} - n_\beta n_\gamma)\right]\left(\frac{\partial u_\sigma}{\partial x_\gamma} + \frac{\partial u_\gamma}{\partial x_\sigma} - \tfrac{2}{3}\delta_{\sigma\gamma}\,\mathrm{div}\,\vec{u}\right), \tag{3.8}$$

where $v_l = 4v/3 + \xi/\rho_0$ and $n_\sigma = k_\sigma/k$. These long-range fluctuations are determined by the stationary nonequilibrium properties of the system and vanish with the vanishing of the flow velocity gradient. The above expressions describe the velocity fields. The full set of one-time CFs of nonequilibrium fluctuations of hydrodynamic fields in a stationary temperature gradient, but without anisotropy tensor, can be found in [4]. The correlators (3.8) and similar ones, found in kinetic gas theory, incorporate nonequilibrium properties which determine gas as a continuum medium. Therefore, the results can be applied for inhomogeneous liquid flows as well.

The long-range correlations due to nonequilibrium thermal fluctuations in an inhomogeneous laminar flow increase their intensity with the increase of the degree of inhomogeneity of flow. However, in a stable flow these intensities are small compared to the correlations in a developed turbulent flow. Nevertheless, these correlations cause weak stochastic mixing in inhomogeneous fluids. These nonequilibrium correlations of thermal nature should anomalously increase at the instability point, forming a structure of turbulent flow or another regime of stable laminar flow. However, the result (3.8) is valid provided that $vk^2 \gg \Gamma$, therefore the influence of the long-range fluctuations on the transfer processes in a weak nonequilibrium fluid is small. Nevertheless these results are important for the asymptotic behaviour of time dependent CFs, which determine the kinetic coefficients in Kubo relations.

We will restrict ourselves to the case of an incompressible liquid. For the Fourier transforms of the velocity and the deviator of the stress tensor one has, after eliminating the pressure in (3.2)

$$\left(\frac{\partial}{\partial t} - \Gamma_{\alpha\lambda} k_\lambda \frac{\partial}{\partial k_\sigma}\right) v_\alpha(k,t) = -A_{\alpha\beta}\Gamma_{\beta\sigma} v_\sigma(k,t) + i v_{12} P_{\alpha\beta} k_\sigma Q_{\beta\sigma}(k,t) - vk^2 v_\alpha(k,t)$$
(3.9)

$$\left(\frac{\partial}{\partial t} - \Gamma_{\alpha\lambda} k_\lambda \frac{\partial}{\partial k_\sigma}\right) Q_{\alpha\beta}(k,t) = -\left(\frac{1}{\tau} + Dk^2\right) Q_{\alpha\beta}(k,t) + i v_{21}(k_\alpha v_\beta(k,t))^s$$
(3.10)

$$\vec{k}\vec{v}(k,t) = 0.$$

Here we have omitted the spontaneous sources and have introduced the tensors

$$A_{\alpha\beta} = \delta_{\alpha\beta} - 2\frac{k_\alpha k_\beta}{k^2}, \quad P_{\alpha\beta} = \delta_{\alpha\beta} - \frac{k_\alpha k_\beta}{k^2},$$
(3.11)

Following [38,21] we introduce time independent wave vectors

$$\frac{\partial k_\sigma}{\partial t} = -\Gamma_{\lambda\sigma} k_\lambda$$

which evolve in time ($\Gamma^T$ is transpose matrix) as

$$k_\sigma(t) = \left(e^{-\Gamma^T t}\right)_{\alpha\lambda} k_\lambda(0).$$
(3.12)

The formal solution of the problem with initial conditions can be found from (3.9)

$$v_\alpha(k,t) = G_{\alpha\sigma}(t) P_{\sigma\lambda}(0) v_\lambda(k(0),0) +$$

$$iv_{12} \int_0^t dt' T_{\alpha\sigma}(t,t') P_{\sigma\beta}(t') k_\lambda(t') Q_{\beta\lambda}(k(t'),t') \qquad (3.13)$$

Where we have introduced the notations

$$G_{\sigma\sigma'}(t) = g_{\sigma\sigma'}(t) e^{-\varphi(t)}, \quad g_{\sigma\sigma'}(t) = \left(\exp\left(-\int_0^t dt' A(t') \cdot \Gamma\right)\right)_{\sigma\sigma'}, \qquad (3.14)$$

$$\varphi(t) = v \int_0^t dt' k^2(t') \qquad (3.15)$$

and the dynamic Oseen tensor

$$T_{\alpha\sigma}(t',t) = \exp(\varphi(t') - \varphi(t)) g_{\alpha\alpha'}(t) g_{\alpha'\beta}^{-1}(t') P_{\beta\sigma}(t') \qquad (3.16)$$

The tensor of transverse projection in (3.16) is

$$P_{\beta\sigma}(t) = \delta_{\beta\sigma} - \frac{k_\beta(t) k_\sigma(t)}{k^2(t)}.$$

The Oseen tensor at $G \neq 0$ depends not only on the difference $t - t'$, which makes the problem much more complex. In a simplest case of a plane Couette flow

$$\Gamma_{\alpha\beta} = \Gamma \delta_{\alpha 1} \delta_{\beta 2} \qquad (3.17)$$

one has for the components of the time dependent wave vector and for the function (3.15)

$$\vec{k}(t) = (k_1, k_2 - \Gamma t k_1, k_3), \qquad (3.18)$$

$$\varphi(t) = v[\tfrac{1}{3} t^3 \Gamma^2 k_1(0) - \Gamma t^2 k_1(0) k_2(0) + t k^2(0)]. \qquad (3.19)$$

The matrix $g(t)$ and inverse one $g^{-1}(t)$ can be found in [38] and are too bulky to write down here.

Substituting (3.13) into the evolution equation for the anisotropy tensor (3.10) and averaging with the initial value, we arrive at a closed set of equations for the CF

$$\Phi_{\alpha\beta\alpha'\beta'}(k,t) = \langle Q_{\alpha\beta}(k,t) Q_{\alpha'\beta'}^*(k,0) \rangle \qquad (3.20)$$

$$\left(\frac{\partial}{\partial t} - \Gamma_{\alpha\lambda} k_\lambda \frac{\partial}{\partial k_\sigma} + \frac{1}{\tau} + Dk^2\right) \Phi_{\alpha\beta\alpha'\beta'}(k,t) =$$

$$- 2b \left(k_\beta(0) \int_0^t dt' T_{\alpha\sigma}(t,t') P_{\sigma\sigma'}(t') k_\lambda(t')\right)^s \langle Q_{\sigma'\lambda}(k(t'),t') Q_{\alpha'\beta'}^*(k(0),0) \rangle \qquad (3.21)$$

Here the coefficient $b$ is the same as in (2.14), and the index $s$ denotes a traceless symmetric (with respect to indices $\alpha\beta$) part of tensor at the right hand side of the equation. In fact, this equation follows from (3.5) and should be accompanied with the initial condition for the CFs (3.6). In the thermodynamic equilibrium at $Ð=0$ these initial CFs of the inertia tensor can be easily found by integrating the spectral density (2.13) over all frequencies Ï

$$\Phi_{\alpha\beta\alpha'\beta'}(k, t=0|\Gamma=0) = \pi\theta \Delta_{\alpha\beta\alpha'\beta'}, \tag{3.22}$$

The result does not depend on fluctuations of the velocity. In the presence of a simple shear flow the CFs

$$\Phi_{\alpha\beta\alpha'\beta'}(k, t=0|\Gamma) = \langle Q_{\alpha\beta}(k,t') Q^*_{\alpha'\beta'}(k,t') \rangle \tag{3.23}$$

can be found, provided that the correlation of fluctuations of the velocity gradient and the anisotropy tensor in a nonequilibrium stationary state neglected. Using the Furutsu-Novikov formula [39,40] for the correlator of Gauss noise and a functional of it (see also [12]), one can find for the stationary CF

$$\left(-\tfrac{1}{2}\Gamma_{\alpha\lambda}k_\lambda \frac{\partial}{\partial k_\sigma} + \frac{1}{\tau} + Dk^2\right)\Phi_{\alpha\beta\alpha'\beta'}(k,0|\Gamma) = \pi\theta\left(\frac{1}{\tau} + Dk^2\right)\Delta_{\alpha\beta\alpha'\beta'}. \tag{3.24}$$

The particular solution of this inhomogeneous equitation is

$$\Phi_{\alpha\beta\alpha'\beta'}(k, t=0|\Gamma) = \pi\theta\Delta_{\alpha\beta\alpha'\beta'}\int_0^\infty dx\left(\frac{1}{\tau} + Dk^2(x)\right)\exp\left(-\int_0^x dy\left(\frac{1}{\tau} + Dk^2(y)\right)\right) = \pi\theta\Delta_{\alpha\beta\alpha'\beta'} \tag{3.25}$$

It coincides with (3.22). We will correct (3.24) by taking into account the velocity approximately.

The equations (3.21) are very complex and an exact solution is hardly possible. We suppose that the tensorial dependence of the CFs (3.20) is the same as in the stationary CFs

$$\Phi_{\alpha\beta\alpha'\beta'}(k,t|\Gamma) = \pi\theta\Delta_{\alpha\beta\alpha'\beta'}\Phi(k,t). \tag{3.26}$$

Substituting this we have for the amplitude $\Phi(k,t)$ an integro-differential equation

$$\frac{\partial}{\partial t}\Phi(k,t) = -\left(\frac{1}{\tau} + Dk^2(t)\right)\Phi(k,t) - \int_0^t dt' N(t,t')\Phi(k,t') \tag{3.27}$$

with kernel

$$N(t,t') = \tfrac{2}{3}bk_\alpha(t)k_\rho(t')T_{\beta\sigma}(t,t')P_{\sigma\lambda}(t')\Delta_{\lambda\rho\alpha\beta}.$$
(3.28)

The kernel is changing much faster than the function $\Phi(k,t)$, therefore we move the CF out of the integral with an upper boundary value. Such an approximation has been used before, e.g. in [41],

for the analysis of CFs of molecular variables. The solution is now simply to find, and it reads [38]

$$\Phi(k,t) = \Phi(k,0)\exp\left[-\int_0^t dt'(\tfrac{1}{\tau}+Dk^2(t'))-Z(t)\right], \quad Z(t) = \int_0^t dt'\int_0^{t'} dt'' N(t',t'')$$

(3.29)

The Oseen tensor in this expression does not allow for explicit integration. Let us consider the asymptotic behaviour at $\Gamma t \gg 1$. Using the explicit expression for the tensor $D\phi$, we have

$$Z(t) \approx \tfrac{4}{3} b\Gamma^2 k_1^2 \int_0^t dt' t' \int_0^{t'} dt'' t'' \exp\left(-\tfrac{1}{3}\nu\Gamma^2 k_1^2 (t'^3 - t''^3)\right)$$

(3.30)

The internal integration gives a degenerate hyperbolic function [36]

$$\int_0^{t'} dt'' t'' \exp\left(\tfrac{1}{3}\nu\Gamma^2 k_1^2 t''^3\right) = \tfrac{1}{2} t'^2 \Phi\left(\tfrac{2}{3},\tfrac{5}{3},\tfrac{1}{3}\nu\Gamma^2 k_1^2 t'^3\right)$$

(3.31)

The second integration can be rewritten as

$$Z(t) \approx \frac{2b\Gamma^2 k_1^2}{15(\Gamma\lambda)^4}[2\Gamma\lambda t + Z_0], \quad \lambda = \left(\frac{\nu k_1^2}{3\Gamma}\right)^{1/3}, \quad Z_0 = \int_0^{(\Gamma\lambda t)^3} x^{1/3}\left[\Phi(1,\tfrac{5}{3},-x)-\frac{2}{3x}\right]dx$$

(3.32)

For large $\Gamma t$, we have

$$Z_0 = \int_0^\infty x^{1/3}\left[\Phi(1,\tfrac{5}{3},-x)-\frac{2}{3x}\right]dx + o\left(\frac{1}{(\Gamma\lambda t)^2}\right)$$

(3.33)

Finally, the amplitude in (3.26) has the form

$$\Phi(k,t) = \Phi(k,0)\exp\left[-\frac{t}{\tau_e} - D(\tfrac{1}{3}t^3\Gamma^2 k_1^2 - t^2\Gamma k_1 k_2 + tk^2) - (r_1|k_1|)^{-3/2}\right],$$

(3.34)

where we have introduced the effective relaxation time and the length parameter

$$\frac{1}{\tau_e} = \frac{1}{\tau} + \frac{4b}{5\nu}, \quad \frac{1}{r_1} = \left(\frac{2bZ_0}{5\nu\Gamma}\right)^{3/4}\left(\frac{3\Gamma}{\nu}\right)^{1/4}$$

(3.35)

Coming back to the equation for the single-time CF (3.23), and using the dynamic equation for the inertia tensor (3.10) and the single-time correlators of random sources $F_{\alpha\beta}$ and the inertia tensor, we find

$$\left\langle Q_{\alpha\beta}\frac{\partial Q^*_{\alpha\beta}}{\partial t} + \frac{\partial Q_{\alpha\beta}}{\partial t}Q^*_{\alpha\beta}\right\rangle - \Gamma_{\alpha\lambda}k_\lambda\frac{\partial}{\partial k_\alpha}\langle Q_{\alpha\beta}Q^*_{\alpha\beta}\rangle = -2(\tfrac{1}{\tau}+Dk^2)\langle Q_{\alpha\beta}Q^*_{\alpha\beta}\rangle -$$

$$iv_{21}\langle Q_{\alpha\beta}(k_\alpha v^*_\beta)\rangle + iv_{21}\langle(k_\alpha v_\beta)'Q^*_{\alpha\beta}\rangle + 2\pi\theta(\tfrac{1}{\tau}+Dk^2)\Delta_{\alpha\beta\alpha'\beta'}$$

(3.36)

Substituting the velocity (3.13) into the r.h.s. of (3.36) and, using the same approximations as for (3.27), we find in the stationary limit $t \to \infty$

$$\left(-\tfrac{1}{2}\Gamma_{\alpha\lambda} k_\lambda \frac{\partial}{\partial k_\sigma} + \frac{1}{\tau_e} + Dk^2\right) \Phi(k,0) = \frac{1}{\tau} + Dk^2, \qquad (3.37)$$

where we have taken into account the limiting value of the additional term in (3.36)

$$\frac{4b}{5\nu} = \lim_{t \to \infty} \frac{\partial Z(t)}{\partial t}.$$

Finally, the amplitude of the single-time CF of the inertia tensor is

$$\Phi(k,0) = 1 - \frac{4b}{5\nu}\int_0^\infty dx \exp\left(-\frac{1}{\tau_e} - D(\tfrac{1}{12} x^3 \Gamma^2 k_1^2 - \tfrac{1}{2} x^2 \Gamma k_1 k_2 + xk^2)\right) \qquad (3.38)$$

It differs from the equilibrium one by an extra term accounting for the velocity fluctuations.

Comparison of (3.34, 3.38) with the corresponding expressions without shear flow makes us conclude, that shear flow changes all parameters of the process, like e.g. the amplitude and the decay coefficient. These quantities depend both on value of G and on the orientation of the wave vector with respect to the flow. Only at the beginning of the process the fluctuating variables have the same value of the decay coefficient compared to the variables in a non-moving fluid. In this case, the decay coefficient decreases for the fluctuations with the wave numbers satisfying $\Gamma k_1 k_2 > 0$, and it increases for the fluctuations for which $\Gamma k_1 k_2 < 0$. At $k_2 \to 0$ the sign of $k_1$ does not influence the asymptotic behaviour of the CF of anisotropy. In case of a zero component of the wave vector in the direction of flow $(k_1 = 0)$, one sees from the equation of motion that fluctuation excitation is not sensitive to the flow and it dissipates the same way as in a fluid at rest.

**4. Space-time dependent CF of inertia tensor**

We can use the results (3.34) to find the main contribution to the CF of fluctuations of the inertia tensor in the presence of shear. For the sake of simplicity, we restrict ourselves only to the first term in the single-time CFs (3.38). In this case we have

$$\Phi(\vec{r},t) = \int d\vec{k}\, e^{i\vec{k}\vec{r}} \Phi(\vec{k},t) = \int d\vec{k}\, \exp\left[i\vec{k}\vec{r} - D(\tfrac{1}{3}t^3\Gamma^2 k_1^2 - t^2\Gamma k_1 k_2 + tk^2) - (r_1|k_1|)^{-3/2} - \frac{t}{\tau_e}\right]. \qquad (4.1)$$

Integration over $k_3$ and $k_2$ is simple and we are left only with the integral over $k_1$

$$\Phi(\vec{r},t) = \frac{\pi}{4Dt} \exp\left(-\frac{t}{\tau_e} - \frac{y^2 + z^2}{4Dt}\right) \int dk_1 \exp\left[-ik_1(x + \tfrac{1}{2}\Gamma ty) - \tfrac{1}{12} a_1^2 k_1^2 - (r_1|k_1|)^{-3/2}\right],$$

$$a_1^2 = Dt(\Gamma^2 t^2 + 12). \qquad (4.2)$$

Making use of the integral

$$\exp(-\tfrac{1}{12}a_1^2 k_1^2) = \sqrt{12\pi} \int dx \exp(-3x^2 + ix a_1 k_1)$$

and introducing the notations $\alpha = \dfrac{a_1}{r_1}$, $\tilde{x} = \dfrac{x + \tfrac{1}{2}\Gamma t}{a_1}$ we rewrite the remaining integration as a convolution

$$\int dk_1 (\cdots) = \frac{\sqrt{12\pi}}{a_1} \int_{-\infty}^{\infty} dx' e^{-3(\tilde{x}-x')^2} g(x') \tag{4.3}$$

where

$$g(x') = 2\alpha^{2/3} \int_0^{\infty} d\zeta \exp\left[-\alpha^{2/3}(ix'\zeta + \zeta^{-2/3})\right]. \tag{4.4}$$

For large $\Gamma t \gg 1$, we have $\alpha \gg 1$. Using the stationary phase method [42], the main contribution to the integral reads

$$g(x) \approx \sqrt{12\pi}\, \alpha (\tfrac{3}{2}\alpha ix)^{-1/4} \exp\left[-\tfrac{5}{3}(\tfrac{3}{2}\alpha ix)^{2/3}\right]. \tag{4.5}$$

After substituting this result into (4.3) and changing the integration variable, we again have an expression with a large exponent

$$\int dk_1(\cdots) = \frac{12\pi}{a_1} \int_{-\infty}^{\infty} d\sigma (\tfrac{3}{2}i\sigma)^{-1/4} \exp\left[-\sqrt{\alpha}\left(3(\sigma - \tilde{x}\alpha^{-1/4})^2 + \tfrac{5}{3}(\tfrac{3}{2}i\sigma)^{-2/3}\right)\right] \tag{4.6}$$

Using the stationary phase method for the new integral, we have for the asymptotic of the CF (4.2)

$$\Phi(\vec{r},t) = \Phi_0(\vec{r},t|\Gamma) \cdot \Phi_1(\vec{r},t|\Gamma) \tag{4.7}$$

with

$$\Phi_0(\vec{r},t|\Gamma) = \left(\frac{\pi}{4Dt}\right)^{3/2} \frac{1}{(1+\tfrac{1}{12}\Gamma^2 t^2)^{1/2}} \exp\left[-\frac{t}{\tau} - \frac{(x+\tfrac{1}{2}\Gamma ty)^2}{4Dt(1+\tfrac{1}{12}\Gamma^2 t^2)} - \frac{y^2+z^2}{4Dt}\right], \tag{4.8}$$

and

$$\Phi_1(\vec{r},t|\Gamma) \propto \operatorname{Re} \exp\left[-\frac{4b}{5v}t - i\frac{x+\tfrac{1}{2}\Gamma ty}{a_1}\left(\frac{8a_1}{r_1}\right)^{1/4} - \left(\frac{8a_1}{9r_1}\right)^{1/2}\right]. \tag{4.9}$$

The function (4.8) is the amplitude of the CF of fluctuations of the density of the inertia tensor without influence of velocity fluctuations. The function (4.9) (written down without an unimportant coefficient) is the asymptotic result of the influence of the velocity fluctuations in the presence of simple steady shear at large $\Gamma t$. The coordinate dependence of these CFs demonstrates the presence of an additional decay in time of fluctuating excitations in the shear plane.

The obtained CFs allows for the limit $\vec{r} \to 0$, which corresponds to the time dependence of the CFs of a Lagrange particle

$$\Phi^L(t) \propto \frac{\theta}{[(Dt)^3 (\Gamma^2 t^2 + 12)]^{1/2}} \exp\left[-\frac{t}{\tau_e} - \left(\frac{64 Dt(\Gamma^2 t^2 + 12)}{81 r_1^2}\right)^{1/4}\right]. \quad (4.10)$$

Neglecting the coefficient and using the expression (3.35) for $r_1$ the second term in the exponent can be rewritten as

$$\left(\frac{b}{\Gamma v}\right)^{3/4} \left(\frac{D}{v} \Gamma t (\Gamma^2 t^2 + 12)\right)^{1/4}. \quad (4.11)$$

This expression is small at any $t$ in the hydrodynamic range. Therefore its influence on the time behaviour of CFs is small.

## 5. Amplitude of stationary CF of fluctuations of inertia tensor

The Fourier transform of the amplitude of the single-time CF of fluctuations of the inertia tensor has been found in the third chapter, and is determined by (3.38) where velocity fluctuations are taken into account as well. We introduce a new notation for the nonequilibrium additional contribution and rewrite it in the form

$$\delta\Phi(\vec{k}, 0) = -\frac{8bk_2}{5\sqrt{\pi} k_1} \exp\left[-\frac{2Dk_2}{\Gamma k_1}\left(k^2 - \tfrac{2}{3}k_3^2 + \frac{1}{\tau_e D}\right)\right] \int_{-1}^{\infty} dy \exp\left[-\frac{2Dk_2^3}{\Gamma k_1}(\tfrac{1}{3}y^3 + \kappa y)\right], \quad (5.1)$$

$$\kappa = \frac{1}{k_2^2}\left(k_1^2 + k_3^2 + \frac{1}{\tau_e D}\right).$$

We split the integration into two contributions over positive and negative values of $y$. The first one gives Lommel function [36]

$$\int_0^\infty dy \exp\left[-\lambda^3(y^3 + 3\kappa y)\right] = \tfrac{2}{3}\sqrt{\kappa}\, S_{0,1/3}(2\lambda^3 \kappa^{3/2}).$$

Therefore, we have

$$\int_{-1}^{\infty} dy (\ldots) = \tfrac{2}{3}\sqrt{\kappa}\, S_{0,1/3}(2\lambda^3 \kappa^{3/2}) + \int_0^1 dy \exp\left[\lambda^3(y^3 + 3\kappa y)\right], \quad \lambda^3 = \frac{2Dk_2^3}{3\Gamma k_1}. \quad (5.2)$$

And again, we are interested in asymptotic behaviour of the CF (5.1) at large $G$, which means

small γ. Expanding (5.2) in series, we have

$$\delta\Phi(\vec{k},0) = -\frac{8bk_2}{5\sqrt{\Gamma}k_1}\exp\left[-\frac{2Dk_2}{\Gamma k_1}\left(k^2 - \tfrac{2}{3}k_3^2 + \frac{1}{\tau_e D}\right)\right]\left[1+o(\lambda^3)\right] \quad (5.3)$$

Let us consider the real space representation

$$\delta\Phi(\vec{r},0) = \int d\vec{k}\, \exp(i\vec{k}\vec{r})\, \delta\Phi(\vec{k},0).$$

The integration over $k_3$ is simple. The integration over $k_1, k_2$ can be rewritten in polar coordinates. Then, the integration over the radial variable gives

$$\delta\Phi(\vec{r},0) = \frac{8b\sqrt{\pi\Gamma}}{5\sqrt{\Gamma}\sqrt{2D}}\int d\hat{k}_1\, d\hat{k}_2 \sqrt{\frac{\hat{k}_2}{\hat{k}_1}}\exp\left(-\zeta\frac{\hat{k}_2}{\hat{k}_1} - \gamma\frac{\hat{k}_1}{\hat{k}_2}\right)\frac{\gamma_1}{\Lambda^2}\Phi(1,\tfrac{1}{2};-\tfrac{1}{4}\gamma_1) \quad (5.4)$$

Here $F(n,m;x)$, is a degenerate hyperbolic function, and the integration is taken over unit vectors $\hat{k}_{1,2} = k_{1,2}/k$ where $\hat{k}_1^2 + \hat{k}_2^2 = 1$, and we have used the following notations

$$\zeta = \frac{1}{\tau_e \Gamma},\ \gamma = \frac{z^2\Gamma}{4D},\ \Lambda = \hat{k}_1 x + \hat{k}_2 y,\ \gamma_1 = \Lambda^2\left[\frac{D\hat{k}_2}{\Gamma\hat{k}_1}(\hat{k}_1^2 + \tfrac{1}{3}\hat{k}_2^2)\right]^{-1} \quad (5.5)$$

Analogously to (5.3) we find instead of (5.4) the following result

$$\delta\Phi(\vec{r},0) \approx \frac{32b\sqrt{\pi\Gamma}}{5\sqrt{\Gamma}\sqrt{D}}\int_0^{\pi/2} d\varphi\left[\frac{\exp(-\zeta\,\mathrm{tg}\varphi - \gamma\,\mathrm{ctg}\varphi)\sqrt{\mathrm{tg}\varphi}}{(x\cos\varphi + y\sin\varphi)^2} + \frac{\exp(-\zeta\,\mathrm{ctg}\varphi - \gamma\,\mathrm{tg}\varphi)\sqrt{\mathrm{ctg}\varphi}}{(x\sin\varphi - y\cos\varphi)^2}\right] \quad (5.6)$$

After changing of integration variable in each term, the integral can be transformed into the expression

$$\int_0^{\pi/2} d\varphi(\ldots) = -\frac{\partial}{\partial y}\int_0^\infty \frac{dt}{\sqrt{t}}\exp\left(-\zeta t - \frac{\gamma}{t}\right)\left(\frac{1}{x+yt} + \frac{1}{x-yt}\right) \quad (5.7)$$

After the expansion of the fraction in series of $yt$ and integration by parts, the integral transforms into the summation over the index of MacDonald function. We make use of the result of [43]

$$\sum_{n=0}^\infty (\mp a)^n K_{n+\tfrac{1}{2}}(w) = \frac{\pi\sqrt{w}}{\sqrt{8}}(1+a^{-1})e^{-w}\Psi(1,\tfrac{3}{2};\tfrac{w}{2}(2\pm a \pm a^{-1})) \quad (5.8)$$

where $\Psi(n,m;x)$ is a degenerate hyperbolic function. Finally, we have arrived at a rigorous result for (5.7). We consider now the case $\gamma \gg 1$ and $\zeta \ll 1$, with the product $\gamma\zeta = z^2/4D\tau_e$. The main contribution to the stationary CF reads

$$\delta\Phi(\vec{r}, 0) = -\frac{4b(16\pi)^{3/2}}{5\sqrt{\Gamma^2}} \frac{x}{(yz)^3} \exp\left[-\sqrt{\frac{z^2}{D\tau_e}}\right]\left[1+o(\frac{1}{\sqrt{\Gamma}})\right]. \qquad (5.9)$$

For small $z$, $|z| << \sqrt{D\tau_e}$, the asymptotic of the CF in real space is a slow decaying function in the directions perpendicular to the flow velocity and an increasing function in the direction of the velocity gradient.

Two-point space-time CFs of fluctuations of the inertia tensor (4.7) should be modified by taking into account the result (5.9)

$$\Phi(\vec{r},t) \to \Phi(\vec{r},t) + \int d\vec{r}' \Phi(\vec{r}-\vec{r}',t)\, \delta\Phi(\vec{r}', 0). \qquad (5.10)$$

In the same way one can estimate the additional contribution to the stationary CF of a Lagrange particle, i.e. at coinciding coordinates of the inertia tensor. This can be done by putting $x=y=z=0$ in (5.4). As a result, we have

$$\delta\Phi^L(\vec{r}=0,0) \propto -\frac{\theta b}{\nu\Gamma} \frac{1}{(D\tau_e)^{3/2}}. \qquad (5.11)$$

______________________________________

In conclusion we mention that we have obtained the correlation functions of the anisotropy tensor in a liquid with anisotropic molecules in equilibrium and nonequilibrium conditions in the presence of steady simple shear, based on generalized hydrodynamics. The asymptotic behaviour of the functions discussed in detail will be used in future work for possible experimental verification.

**Acknowledgements.**

The very idea of this work originated from discussions with Dick Bedeaux, and especially after reading a yet unpublished Bedeaux and Rubi's paper [28].

Alexander Zatovsky thanks NWO (Nederlandse Organisatie voor Wetenschappelijk Onderzoek) for a grant, which enabled him to visit Dick Bedeaux's group at the Leiden University.

Authors like to thank Mathieu Ernst for his comments about this work. Reading of the manuscript by Agur Sevink is thankfully appreciated.